\documentclass[twocolumn,aps,pra,amsmath,amssymb,10pt]{revtex4-1}
\usepackage[latin9]{inputenc}
\usepackage{color}
\usepackage{graphicx}

\usepackage{dcolumn}\DeclareMathOperator\erfc{erfc}

\newcommand{\bitem}{\begin{itemize}}
\newcommand{\fitem}{\end{itemize}}
\newcommand{\beq}{\begin{equation}}
\newcommand{\eeq}{\end{equation}}
\newcommand{\beqa}{\begin{eqnarray}}
\newcommand{\eeqa}{\end{eqnarray}}

\begin{document}

\title{The Atomic Lighthouse Effect}

\author{C. E. M\'aximo$^{1}$, R. Kaiser$^{2}$, Ph.W. Courteille$^{1}$ and
R. Bachelard$^{1}$}

\affiliation{ $^{1}$Instituto de F\'{i}sica de S\~ao Carlos, Universidade de S\~ao
Paulo,13560-970 S\~ao Carlos, SP, Brazil\\
 $^{2}$Universit\'e de Nice Sophia Antipolis, CNRS, Institut Non-Lin\'eaire
de Nice, UMR 7335, F-06560 Valbonne, France}

\date{\today}
\begin{abstract}
We investigate the deflection of light by a cold atomic cloud when
the light-matter interaction is locally tuned via the Zeeman effect
using magnetic field gradients. This ``lighthouse'' effect is strongest
in the single-scattering regime, where deviation of the incident field
is largest. For optically dense samples, the deviation is reduced by
collective effects, as the increase in linewidth leads to a decrease of the magnetic field efficiency.
\end{abstract}
\maketitle

\section{Introduction}

The interference of light scattered by different particles of an ensemble
is at the origin of a variety of collective phenomena such as superradiance~\cite{Dicke1954},
Bragg scattering or collective frequency shifts~\cite{Friedberg1973,Rohlsberger2010,Keaveney2012}.
The phenomena can be classified in two distinct regimes according
to whether the scatterers interact with each other via mediation of
the incident light, or not. Bragg scattering, for example, is the
result of a far-field interference of the light waves scattered by
an optically dilute, periodic structure. In this case, even in the
absence of communication between the scatterers, the radiated light
pattern provides information on the scattering structure, a fact that
is extensively used,e.g.~in crystallography. On the other hand, optically
dense structures lead to multiple scattering and strong interference
in the near-field (i.e.~within the structure) between the light waves
scattered from different particles. One example is the opening of
photonic band gaps in period structures~\cite{Deutsch1995,Antezza2009,Schilke2011,Antezza2013,Samoylova2014}.

Cold atomic clouds are particularly attractive experimentation platforms
as powerful techniques not only allow to shape the density distribution
but also to fine-tune the light-matter interaction over wide ranges.
Here we show that sufficiently cold atomic clouds exposed to a gradient
of the strength of the light-matter interaction deflect light due
to collective scattering in the single-scattering regime. We propose to implement the required gradient
by an inhomogeneous magnetic field exploiting the Zeeman
effect. Because this phenomenon is reminiscent of an effect studied in
nuclear physics called ``Lighthouse effect''~\cite{Rohlsberger2000,Rohlsberger2001,Rohlsberger2001a,Roth2005},
we call this effect the ``atomic lighthouse effect''. 

The nuclear lighthouse effect was used to perform spectroscopy and the 
transformation from time to angular coordinates allowed to detect timescales difficult to achieve with present detection schemes. 
Similar light deviation
effects have been experimentally observed for light passing through an atomic vapour in slow light and electromagnetically induced transparency schemes using either magnetically 
~\cite{Karpa2006, Weitz2010} or optically induced gradients~\cite{Bloch2009, Sautenkov2010}.

We will show in this paper that the atomic lighthouse effect can be obtained on a two level scheme and is 
a result of the interference of the light radiated by independent
atoms. Similarly to Bragg scattering, it is thus fully determined
by the single-photon structure factor of the atomic cloud. However,
light-induced interatomic cooperation dramatically \emph{alters}
the lighthouse effect in the case of optically dense samples. We prove
this via calculations and simulations accounting for the light-induced
interactions between the atoms in the multiple scattering regime.
The alteration can be understood as an increase in the atomic linewidth, due to the atoms cooperation, that reduces the Zeeman effect. Hence, the lighthouse reduction provides a direct signature of cooperativity.

\section{Coupled dipole model}

We describe the light deviation by a cloud of atoms with a model treating
all atomic dipoles as being coupled via the incident light. This allows
us to account for interferences between these dipoles in the optically
dense regime. Photon random walk approaches are not
sufficient. Furthermore, all dipoles are exposed to a locally varying
atom-light interaction inducing an inhomogeneous phase profile of
the dipole excitation across the atomic cloud.

\begin{figure}
\centering \includegraphics[width=0.9\linewidth]{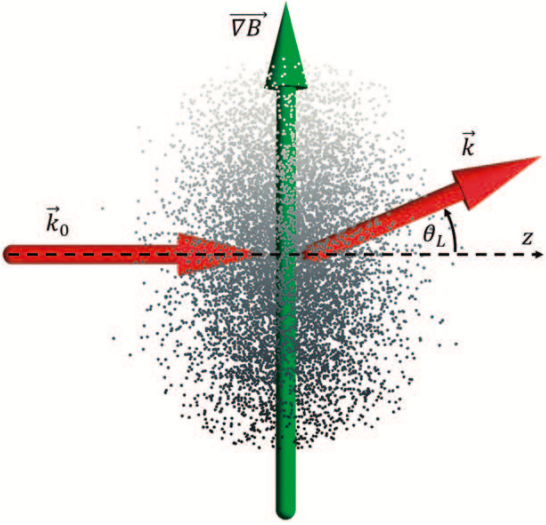} \caption{\label{fig:SetUp} (Color online) Scheme of the atomic lighthouse
effect: An incident laser beam labeled by its wave vector $\mathbf{k}_{0}$,
is scattered from a cloud of cold atoms to which a transverse magnetic
field gradient, $\nabla B$, is applied. As a consequence
the light is deviated by an angle $\theta_{L}$.}
\end{figure}

Let us start by considering an ensemble of $N$ two-level ($g$ and
$e$) atoms, each at a position $\mathbf{r}_{j}$ ($j=1,2,\cdots N$),
driven by a uniform laser beam with wave vector $\mathbf{k}_{0}=k_{0}\mathbf{\hat{z}}$.
The detuning of the light from the atomic resonance is $\Delta_{0}=\omega_{0}-\omega_{a}$,
and the Rabi frequency is $\Omega_{0}=dE_{0}/\hbar$, with $d$ the
dipole matrix element, $E_{0}$ the field amplitude and $\hbar$ the
Planck constant. The atom-light interaction is locally tuned with
a inhomogeneous magnetic field in the quantization direction $\mathbf{B}(\mathbf{r}_{j})=\mathbf{\hat{z}}B(\mathbf{r}_{j})$
(see Fig.\ref{fig:SetUp}).

For the sake of simplicity we consider an electronic transition with
a structureless ground state and single excited state Zeeman level.
This system is described by the following Hamiltonian~\cite{Bienaime2011}
in the rotating wave approximation 
\begin{eqnarray}
\hat{H} & = & \frac{\hbar\Omega_{0}}{2}\sum_{j=1}^{N}\left(\hat{\sigma}_{-}^{(j)}e^{i\Delta_{0}t-i\mathbf{k}_{0}\cdot\mathbf{r}_{j}}+h.c.\right)\nonumber \\
 &  & +\hbar\sum_{j=1}^{N}\sum_{\mathbf{k}}g_{\mathbf{k}}\left(\hat{\sigma}_{-}^{(j)}\hat{a}_{\mathbf{k}}^{\dagger}e^{i\Delta_{\mathbf{k}}t-i\mathbf{k}\cdot\mathbf{r}_{j}}+h.c\right)\nonumber \\
 &  & -\hbar\sum_{j=1}^{N}\Delta(\mathbf{r}_{j})\sigma_{z}^{(j)}.\label{hamiltonian}
\end{eqnarray}
$\Delta_{\mathbf{k}}=\omega_{\mathbf{k}}-\omega_{a}$ is the detuning
between the field emitted into mode $\mathbf{k}$ and the atomic transition
$\omega_{a}$. $g_{\mathbf{k}}=d\sqrt{\omega_{k}/\hbar\epsilon_{0}V_{\nu}}$
is the single-photon Rabi frequency for a photon volume $V_{\nu}$.
$\Delta(\mathbf{r}_{j})=\mu_{B}m_{j}B(\mathbf{r}_{j})/\hbar$ describes
the (inhomogeneous) Zeeman effect, $\hat{\sigma}_{-}^{(j)}=|g_{j}\rangle\langle e_{j}|$
and $\hat{\sigma}_{z}^{(j)}=|e_{j}\rangle\langle e_{j}|-|g_{j}\rangle\langle g_{j}|$
are the Pauli matrices describing the de-excitation and the excited
state population of atom $j$, whereas $\hat{a}_{\mathbf{k}}$ describes
the photon annihilation in the mode $\mathbf{k}$.

Under the Markov approximation and in the linear optics regime a scattering
equation for the dipole excitation $\beta_{j}$ can be derived~\cite{Bienaime2011}: 
\begin{eqnarray}
\frac{d\beta_{j}}{dt} & = & \left[i(\Delta(\mathbf{r}_{j})+\Delta_{0})-\frac{\Gamma}{2}\right]\beta_{j}-\frac{i\Omega_{0}}{2}e^{i\mathbf{k}_{0}\cdot.\mathbf{r}_{j}}\nonumber \\
 &  & -\frac{\Gamma}{2}\sum_{m\neq j}\frac{\exp(ik_{0}|\mathbf{r}_{j}-\mathbf{r}_{m}|)}{ik_{0}|\mathbf{r}_{j}-\mathbf{r}_{m}|}\beta_{m},\label{eq:CD}
\end{eqnarray}
where $\Gamma=V_{\nu}g_{k_{0}}^{2}k_{0}^{2}/\pi c$ the transition
linewidth. Since the laser detuning $\Delta_{0}$ is equivalent to
a magnetic field offset, we may set $\Delta_{0}=0$ without loss of
generality~%
\footnote{Note that this is only true under the assumption of a single excited
state Zeeman level.%
}.

As the lighthouse effect relies on a global phase contrast in the
dipole field rather than on disorder effects, we adopt a fluid description
of the system introducing the local density $\rho(\mathbf{r})$ and
dipole field $\beta(\mathbf{r})$. Then, in the steady state regime,
Eq.~\eqref{eq:CD} turns into a Fredholm equation of the second type
\begin{equation}
\beta(\mathbf{r})=\beta^{(1)}(\mathbf{r})+\int\mbox{d}^{3}\mathbf{r'}\rho(\mathbf{r})K(\mathbf{r},\mathbf{r}')\beta(\mathbf{r}'),\label{fredholm}
\end{equation}
where we have introduced the single-scattering dipole excitation 
\begin{equation}
\beta^{(1)}(\mathbf{r})=\frac{\Omega_{0}}{\Gamma}\frac{e^{i\mathbf{k}_{0}\cdot\mathbf{r}}}{i+2\Delta(\mathbf{r})/\Gamma},\label{eq:SS}
\end{equation}
and the scattering kernel 
\begin{equation}
K(\mathbf{r},\mathbf{r}')=\frac{1}{2i\Delta(\mathbf{r})/\Gamma-1}\frac{e^{ik_{0}|\mathbf{r}-\mathbf{r}'|}}{ik_{0}|\mathbf{r}-\mathbf{r}'|}.
\end{equation}
The consequence of a differential Zeeman effect is intuitive in the single-scattering
limit~(\ref{eq:SS}), where the field $\Delta(\mathbf{r})$ modulates
spatially the phase (and amplitude) of the dipole field, thus modifying
the direction of superradiant emission of the cloud. In the multiple
scattering regime, a more detailed study must be performed in order
to understand how the rescattering of the photons alters this new
phase profile (see Sec.~\ref{sec:MS}).

\section{Single-scattering regime}

For the sake of simplicity, we focus on a linear magnetic field, which
is orthogonal to the laser beam and cancels at the cloud's center:
$B(\mathbf{r})=bx$. We found the lighthouse effect
to be strongest is this geometry. Indeed, for a central detuning $\Delta(\mathbf{r}=\mathbf{0})$ much larger than the one created by the gradient of magnetic field $bR$ ($R$ the cloud radius), the resulting gradient of phase is tuned down by a factor $1/(1+4\Delta(\mathbf{0})^2/\Gamma^2)$, thus reducing the deflection angle.

We describe the cloud's density
distribution by Gaussian spheres, which presents the analytical advantage
of having a factorisable density in Cartesian coordinates: 
\begin{equation}
\rho(x,y,z)=\frac{N}{\left(2\pi\right)^{3/2}R^{3}}e^{-\frac{x^{2}+y^{2}+z^{2}}{2R^{2}}}.
\end{equation}
. The far-field radiated electric field is given by: 
\begin{equation}
E(\mathbf{k})=\frac{\hbar\Gamma }{id}\frac{e^{ik_{0}r}}{r}\int\mbox{d}^{3}\mathbf{r}\rho(\mathbf{r})\beta(\mathbf{r})e^{-i\mathbf{k}\cdot\mathbf{r}},\label{eq:sk}
\end{equation}
where, for the sake of clarity, we have omitted the time-dependent oscillating term $e^{-ik_0ct}$.
The field resulting from single scattering $E^{(1)}$ is then calculated
by replacing the complete dipole excitation $\beta$ in Eq.~\eqref{eq:sk}
by the single-scattering excitation $\beta^{(1)}$ {[}see Eq.~\eqref{eq:SS}{]}.
For a Gaussian cloud, introducing the normalized gradient $\alpha=2\mu_{B}b/(\hbar k_{0}\Gamma)$
and the normalized cloud size $\sigma=k_{0}R$, it reads (see Appendix
\ref{app:SDContrib}) \begin{widetext} 
\begin{equation}
E^{(1)}(\mathbf{k})=-\sqrt{\frac{\pi}{2}}\frac{NE_{0}}{\left|\alpha\right|\sigma } \frac{e^{ik_{0}r}}{r}\exp\left(\frac{1}{2\alpha^{2}\sigma^{2}}+\frac{\sin\theta}{\alpha}-\frac{\sigma^{2}}{2}(1-\cos\theta)^{2}\right)\erfc\left(\frac{1}{\left|\alpha\right|\sigma\sqrt{2}}+\frac{\mbox{sign}\left(\alpha\right)\sigma\sin\theta}{\sqrt{2}}\right),\label{eq:skSS}
\end{equation}
\end{widetext} with $\alpha\neq0$ and $0\leq\theta\leq2\pi$.

The expansion in scattering orders is also investigated numerically,
using Gaussian distributions $\{\mathbf{r}_{i}\}$ for direct simulations
of the many-body problem. Then, the single scattering contribution
is obtained by considering the single scattering dipole excitation
$\beta_{j}^{(1)}=(\Omega_{0}/\Gamma)/(i+2\Delta(\mathbf{r}_{j})/\Gamma)$,
from which the radiated field is derived. The following scattering
order is obtained as $\beta^{(2)}_j=\sum_m\mathbf{K}_{jm}\beta^{(1)}_m$, where the
scattering matrix $\mathbf{K}$ has components $\mathbf{K}_{jm}=K(\mathbf{r}_{j},\mathbf{r}_{m})$
for $j\neq m$, and $\mathbf{K}_{jj}=0$. Each higher scattering order
is obtained by applying $K$ on the previous one.

\begin{figure}
\centering \includegraphics[width=1\linewidth]{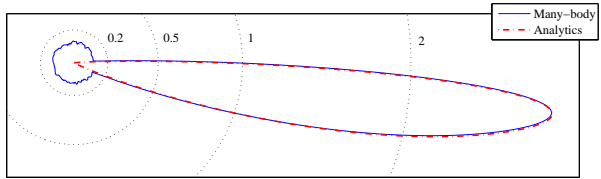} \caption{(Color online) Radiation pattern of light scattered ($|E|^2$) by an atom cloud
under the influence of a differential Zeeman effect. Incident from the
left, the light is deflected by a transverse magnetic field gradient.
The analytical curve (red and dashed) corresponds to Eq.~\eqref{eq:skSS},
whereas the many-body curve (plain and blue) is obtained from the stationary solution
of Eq.~\eqref{eq:CD}. The simulations are made for a cloud of $N=100$
atoms with radius $\sigma=35$ ($b_{0}=0.16$) and $\alpha=1$. The
many-body simulations are averaged over $500$ realizations.}

\label{fig:DiagEm} 
\end{figure}

An illustration of Eq.~\eqref{eq:skSS} exhibiting the lighthouse
effect is shown in Fig.~\ref{fig:DiagEm}. It clearly demonstrates
how the transverse phase gradient deflects the incoming light. Furthermore,
Eq.~\eqref{eq:skSS} appears in excellent agreement with numerical
simulations of the many-body problem~\eqref{eq:CD} realized in a
regime where the cloud is optically thin ($b_{0}=0.16$). Indeed,
the only difference between the two radiation patterns is an (apart
from some noise) isotropic background present in the many-body simulations.
This background has its origin in the atomic disorder, which is naturally
absent from our analytical approach. 

\begin{figure}
\centering\includegraphics[width=1\linewidth]{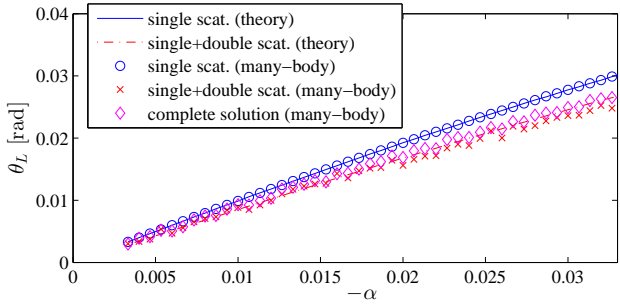} \caption{(Color online) Deflection angle $\theta_{L}$ as a function of the
scaled magnetic field gradient $\alpha$. The curves correspond respectively
to single- (plain blue) and double-scattering (dashed red) contributions.
The circles, crosses and squares stand, respectively, for many-body
single-, double- and full scattering solutions. The simulations were
realized for a cloud with radius $\sigma=21$($b_{0}=0.5)$ of $N=113$
atoms, and averaged over $100$ realizations.}
\label{fig:ScanAlpha} 
\end{figure}

The deflection angle $\theta_{L}$ is obtained by maximizing \eqref{eq:skSS}
over $\theta$, which leads to:\begin{widetext} 
\begin{equation}
\frac{1}{\alpha}-\sigma\sqrt{\frac{2}{\pi}}\frac{\exp\left[-\frac{1}{2}\left(\frac{1}{\left|\alpha\right|\sigma}+\mbox{sign}\left(\alpha\right)\sigma\sin\theta_{L}\right)^{2}\right]}{\erfc\left(\frac{1}{\sqrt{2}}\left(\frac{1}{\left|\alpha\right|\sigma}+\mbox{sign}\left(\alpha\right)\sigma\sin\theta_{L}\right)\right)}=\sigma^{2}\tan\theta_{L}(1-\cos\theta_{L})\label{eq:devang}
\end{equation}
\end{widetext} 
Fig.~\ref{fig:ScanAlpha} compares the calculated
deflection angle to many-body simulations of Eq.~\ref{eq:CD}, confirming
the validity of the above formula.
Under the approximation of a small total detuning over the cloud ($\alpha\sigma\ll 1$), we get
\begin{equation}
\theta_{L}=\sin^{-1}\left(\frac{1-\sqrt{1+4\alpha^{2}\sigma^{2}}}{2\alpha\sigma^{2}}\right)\approx-\alpha.
\end{equation}
In the other limit of large $\alpha$, note that the deviation angle is not limited by the diffraction limit from the cloud size. Indeed, an increasing gradient of magnetic field makes that only a smaller volume of atoms scatter the light efficiently, thus reducing the effective size of the macroscopic scatterer (see Fig.\ref{fig7}).

\begin{figure}
\centering \includegraphics[width=1\linewidth]{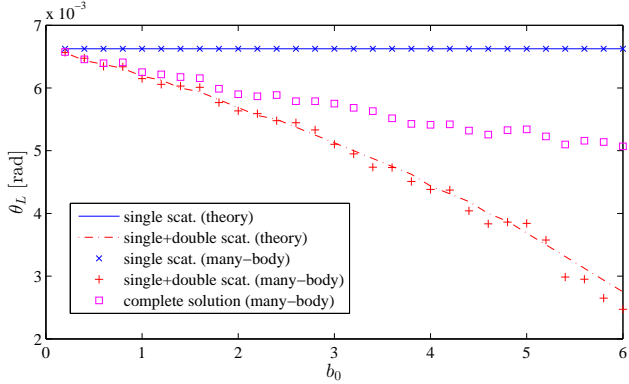} \caption{ (Color online) Deflection angle $\theta_{L}$ as a function of the
optical thickness $b_{0}$. The curves correspond respectively to
single- (plain blue) and double-scattering (dashed red) contributions.
The circles, crosses and squares stand for, respectively, many-body
single-, double- and full scattering solutions. Simulations realized
for a cloud of $N=113$ atoms and radius $\sigma=21$, and averaged
over $100$ realizations.}

\label{fig:ScanB0} 
\end{figure}

Yet, as one increases the optical thickness of the cloud $b_{0}=2N/\sigma^{2}$
(see Fig.~\ref{fig:ScanB0}), the single-scattering prediction loses
its accuracy, pointing to the fact that the optically dense regime
is ruled by photon rescattering. While an exact solution of the three-dimensional
scattering problem including interference does not, to the best of
our knowledge, exist, it is possible to probe the dense regime using,
for example, a multiple scattering expansion~\cite{Ishimaru1978,Lagendijk1988,Nieuwenhuizen1992,Rouabah2014}.
As we will see now, the double-scattering contribution can be evaluated,
providing valuable hints of the dense regime. 

\section{Multiple scattering regime\label{sec:MS}}

An analytical treatment of the multiple scattering regime involves
solving the problem of $N$ fully coupled atom dipoles. Mean-field
approaches such as the timed Dicke state~\cite{Scully2006}, or random
walk approaches neglecting phase coherences, obviously cannot capture
the modification of the atomic phase field. We thus resort to a multiple
scattering expansion, an approach that has proved particularly useful
in the treatment of, e.g., coherent backscattering~\cite{Akkermans1986,Wolf1985,Albada1985,Lagendijk1988}.

The double scattering contribution to the field is obtained in Appendix~\ref{app:SDContrib}
as \begin{widetext} 
\begin{eqnarray}
E^{(2)}(\theta) & = & NE_{0}\frac{\sqrt{\pi}e^{1/(4\alpha^{2}\sigma^{2})+\sin\theta/(2\alpha)-\sigma^{2}(1-\cos\theta)^{2}/4}}{16\left|\alpha\right|\sigma}\frac{e^{ik_{0}r}}{r}\erfc\left(\frac{1}{2\left|\alpha\right|\sigma}+\frac{\mbox{sign}\left(\alpha\right)\sigma\sin\theta}{2}\right)\nonumber \\
 &  & \times\frac{b_{0}}{\cos\frac{\theta}{2}}\left[e^{-\sigma^{2}(1+\sin\theta)^{2}}\erfc\left(-i\sigma(1+\sin\theta)\right)-e^{-\sigma^{2}(1-\sin\theta)^{2}}\erfc\left(-i\sigma(1-\sin\theta)\right)\right].\label{eq:S2f}
\end{eqnarray}
\end{widetext} In Eq.~\eqref{eq:S2f}, the first line describes
the light deviation, while the second line yields the double scattering
process: the prefactor containing the optical thickness is typical
for such an expansion. The validity of this expansion is delimited by
the condition $b_{0}<1$, when single and double scattering are the
dominant contributions.

\begin{figure}
\centering \includegraphics[width=1\linewidth]{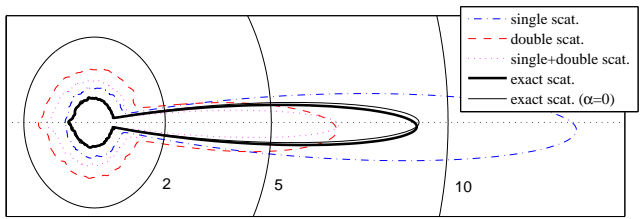} \caption{(Color online) Radiation pattern $|E|^2$ for different scattering orders (see
legend), for the many-body system.The plain thick and thin line correspond respectively to the exact scattering solution with and without gradient of magnetic field. Simulations realized for a cloud
of $N=225$ atoms, $\sigma=21$ ($b_{0}=1$) and $\alpha=0.013$,
and averaged over $500$ configurations. }

\label{fig:DiagEmS1S2} 
\end{figure}

In Fig.~\ref{fig:DiagEmS1S2}, the first, second and complete scattering
solutions for the many-body problem are plotted and compared to the
analytical solution. The single and double scattering contributions
are correctly predicted by the theory. Furthermore, the figure shows
a reduction of the deflection angle as the optical thickness increases.
Thus the rescattering of the light by the atoms tends to erase the
dipole phase gradient imposed by the magnetic field. 

We note that the single and double scattering fields interfere destructively,
since their electric fields have opposite signs. Thus a careful treatment
of the light field amplitude rather than the intensity is crucial,
as already noted in~\cite{Rouabah2014}.

As the optical thickness increases beyond unity, the exact scattering
solution, which contains all scattering orders, deviates from the
prediction obtained for double scattering. This is confirmed by the
analysis performed in Fig.~\ref{fig:ScanB0}, where the deviation
angle appears correctly predicted by the single scattering approach
for $b_{0}\ll1$, and by the double scattering calculations for $b_{0}\leq1$.
The extremum equation for the deflection angle $\theta_{L}$ including
double scattering is straightforwardly derived from Eq.~\ref{eq:S2f}, yet cumbersome, for
which reason it is not presented here.

Features of the deep multiple scattering regime can be captured using modified timed Dicke state: this mean-field ansatz assumes a perfect synchronization of all atomic dipoles~\cite{Scully2006} and, as a consequence, a broadening of the atomic linewidth. For a Gaussian cloud, the collective linewidth is $\Gamma_N=\Gamma(1+b_0/8)$~\cite{Courteille2010}. In our case, we also need to account for an effective reduction of the optical thickness due to the inhomogeneous detuning imposed by the gradient of magnetic field. The average detuning of the atoms compared to those on the optical axis being $\delta_{LH}=\alpha\sigma$, the effective optical thickness for the cloud is $b^{eff}=b_0/(1+4\delta_{LH}^2)$.

\begin{figure}
\centering \includegraphics[width=1\linewidth]{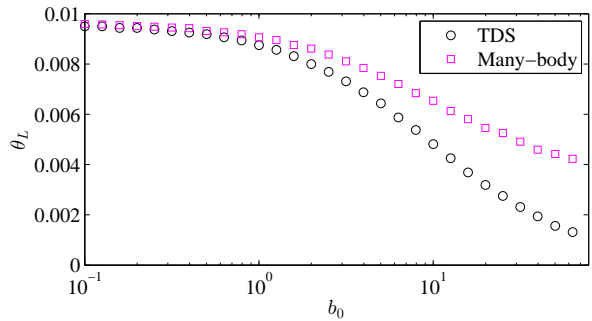} \caption{(Color online) Deviation angle for optically thick samples. Simulations
realized for a cloud of size $\sigma=21$, the number of particles
being determined by the optical thickness, and $\alpha=-0.01$.}
\label{fig:LargeB0} 
\end{figure}

Replacing the atomic linewidth by the collective one $\Gamma(1+b^{eff}/8)$ in the single-scattering prediction, we predict a reduction of the lighthouse effect as the optical density increases: the increase in linewidth leads to a decrease in the normalized gradient $\alpha$. Simulations of the many-body problem confirm that result, and show a qualitative agreement with the prediction of the modified timed Dicke ansatz (see Fig.~\ref{fig:LargeB0}).

\section{Discussion \& experimental perspectives}

We investigated the deflection of light by an atomic cloud under the
influence of a differential Zeeman shift and found that the single-scattering
regime yields maximum deflection. Calculations and simulations for
clouds of larger optical density, which are dominated by multiple
scattering, revealed a reduction of the effect. This observation points
to the fact that the lighthouse deflection is clearly not a cooperative
effect in the sense that it does not result from light-mediated interaction
of the atoms. It is rather similar to Bragg scattering, where the
scattering pattern generated by interference of the radiated waves
provides information on the atomic cloud's structure. As soon as the
atoms interact through their radiation, the emergence of multiple
scattering washes out the phase gradient imposed externally by the
magnetic field.

Our predictions can be experimentally verified, e.g.~with a cloud
of spatially confined cold atoms exposed to a uniform magnetic field
gradient. Measurable deflections are expected when the parameter $\alpha$
is not too small compared to $1$. According to the definition of
$\alpha$ {[}below Eq.~(\ref{eq:sk}){]}, for a typical magnetic
field gradient of $b=100~$G/cm, this requires rather small linewidths
of the atomic transition. A possible system would be a cloud of ultracold
strontium driven on its intercombination line at $\lambda=2\pi/k=689~$nm.
The transition linewidth being $\Gamma=(2\pi)~7.6~$kHz one could
reach $\alpha=0.8$ with the specified gradient.

In our derivations we assumed a single excited state Zeeman level.
However, if the atomic cloud is trapped in a magneto-optical trap
(MOT) we would rather have a distribution of atoms in all Zeeman levels.
Furthermore, the magnetic field does not have the geometry of a uniform
gradient but of a quadrupolar field. These problems, can, however,
be circumvented by suddenly applying a magnetic field offset $B_{0}$
and simultaneously detuning the probe light beam to the Zeeman-shifted
resonance. In that way, one spectrally filters out a single Zeeman
state, i.e.~the probe light predominantly interacts with atoms occupying
a single Zeeman level. This requires $B_{0}\gg\Gamma$, which is easy
to satisfy in the case of a narrow linewidth $\Gamma$.

Typical values for a strontium MOT operated on the narrow intercombination
line are an atom number of $N=2\cdot10^{7}$ and a radial size of
$\bar{R}=70~\mu\text{m}$ (or $\sigma=k\bar{R}\simeq640$) \cite{Chaneliere2004},
giving an optical density of $b_{0}=2N/(k\bar{R})^{2}\simeq100$ for
the intercombination line. In practice, however, narrow transitions
are generally dominated by Doppler broadening. At a temperature of
$4~\mu$K, for example, the Doppler broadening of the intercombination
line is: $k\bar{v}=k\sqrt{k_{B}T/m}=(2\pi)~28~$kHz. Even if the temperature
is at the Doppler limit of the intercombination line, $T_{D}=\hbar\Gamma/k_{B}=365~$nK,
the Doppler width still is $8.5~$kHz.

To prevent blurring of the lighthouse deflection by the thermal atomic
motion, it is important that the Doppler shift be smaller than the
Zeeman shift, $k\bar{v}\ll\bar{R}\partial_{r}B$. In this case, the
main effect of the thermal motion is to reduce the optical thickness
of the atomic cloud, by a factor corresponding to the spectral overlap
between the natural linewidth and the Doppler broadened width, $b_{D}=b_{0}\Gamma/kv$.

\begin{figure}
\centering \includegraphics[width=0.9\linewidth]{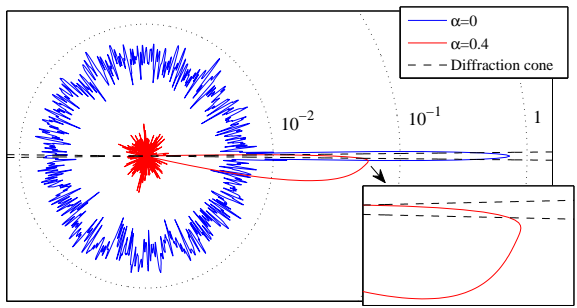} \caption{\label{fig7} (Color online) Radiation pattern normalized by the incident intensity $|E/E_0|^2$ for the many-body system, with (thick red) and without (thin blue) gradient of magnetic field. The inset allows to observe that the lighthouse angle is not diffraction-limited (cone of black dashed lines; see main text). Note that the strong background radiation is due to the limited number of particles that can be simulated. The reduced radiated intensity in presence of the gradient of magnetic field is due to the fact that many atoms are driven out of resonance by it, so their coupling with the light is reduced. Simulations realized for a cloud of $N=15000$ atoms, $\sigma=101$ ($b_{0}=2.94$) and $\alpha=0.4$,
and averaged over $10$ configurations.}
\end{figure}

Another point ruling the detectability of the lighthouse deflection
is, whether it exceeds the probe beam divergence angle. Assuming that
it is optimally matched to the size of the cloud, $w_{0}=\bar{R}=70~\mu$m,
we obtain the divergence angle $\alpha_{div}=\lambda/\pi w_{0}=0.18^{\circ}$.
Since $\sigma=70\mu m/689nm\approx101$ and $\alpha\approx0.4$, the
deflection angle predicted by Eq.\eqref{eq:devang} is $\theta_{L}=1.34^{\circ}$,
so that the deviated beam is easily separated from the incident laser (see Fig.~\ref{fig7}).

Even on regular MOTs operating on strong transitions, the atomic lighthouse effect
should be detectable. Considering, for instance, the broad $^1S_0-^1P_1$ transition 
in strontium ($\Gamma=(2\pi)~32~$MHz), we estimate a deflection angle of $\theta_{L}=0.026^{\circ}$,
which is on the same order of magnitude as the divergence angle, which for a MOT radius of 
typically $\bar R=3~$mm, would be $\alpha_{div}=0.028^\circ$.

\bigskip

The experimental verification of the lighthouse effect would represent
a nice confirmation that our actual understanding of how light is
cooperative scattered by ensembles of particles is correct, as the reduction of the deviation angle provides a direct measure of the collective atomic linewidth.

\section{Acknowledgements}

We acknowledge financial support from IRSES project COSCALI, from USP/COFECUB (projet Uc Ph 123/11) and from GDRI ``Nanomagnetism, Spin Electronics, Quantum Optics and Quantum Technologies''. C.E.M\'aximo, Ph. W. C and R. B. acknowledge support from the Brazilian FAPESP and CNPq agencies.

\appendix
\begin{widetext}

\section{Single- and double-scattering structure factor\label{app:SDContrib}}

In the far-field limit and at a distance $r$, the single scattering field is given by: 
\[
E^{(1)}(\mathbf{k})=-\frac{NE_{0}}{\left(2\pi\right)^{\frac{3}{2}}R^{3}}\frac{e^{ik_0r}}{r}\int_{-\infty}^{\infty}e^{i(1-\cos\theta)k_{0}z}e^{-z^{2}/2R^{2}}dz\int_{-\infty}^{\infty}e^{-i\sin\theta\sin\phi k_{0}y}e^{-y^{2}/2R^{2}}dy\int_{-\infty}^{\infty}\frac{e^{-i\sin\theta\cos\phi k_{0}x}e^{-x^{2}/2R^{2}}}{1+i\alpha k_{0}x}dx.
\]
The first integral gives $\exp(-(1-\cos\theta)^{2}\sigma^{2}/2)$
that yields the forward emission of Rayleigh scattering, in a cone
of width $\sim1/\sigma$. The second integral produces a $\exp(-\sin^{2}\theta\sin^{2}\phi\sigma^{2}/2)$,
yet since the problem is symmetric with respect to the $(\hat{x},\hat{z})$
plane, we restrict ourselves to this plane, taking $\phi=0,\pi$.
Finally, the integral over $x$, that contains the deviation is
\[
\int_{-\infty}^{\infty}\frac{e^{\pm i\sin\theta k_{0}x}e^{-x^{2}/2R^{2}}}{1+i\alpha k_{0}x}dx=\gamma^{2}\int_{-\infty}^{\infty}e^{-x^{2}/2R^{2}}\cos ax\frac{dx}{\gamma^{2}+x^{2}}\mp\mbox{sign}\left(\gamma\right)\left|\gamma\right|\int_{-\infty}^{\infty}e^{-x^{2}/2R^{2}}\sin ax\frac{xdx}{\gamma^{2}+x^{2}}
\]
with $a=k_{0}\sin\theta$ and $\gamma=1/\alpha k_{0}$, and where
the $\pm$ signs refers to the cases $\phi=0,\pi$. Both integrals are solved
using integral formula~\cite{gradshteyn2007} 
\begin{eqnarray}
\int_{-\infty}^{\infty}e^{-x^{2}/2R^{2}}\sin ax\frac{xdx}{\gamma^{2}+x^{2}} & = & -\frac{\pi}{2}e^{\gamma^{2}/2R^{2}}\left[e^{a\left|\gamma\right|}\erfc\left(\frac{1}{\sqrt{2}}\left(\frac{\left|\gamma\right|}{R}+aR\right)\right)-e^{-a\left|\gamma\right|}\erfc\left(\frac{1}{\sqrt{2}}\left(\frac{\left|\gamma\right|}{R}-aR\right)\right)\right],\\
\int_{-\infty}^{\infty}e^{-x^{2}/2R^{2}}\cos ax\frac{dx}{\gamma^{2}+x^{2}} & = & \frac{\pi}{2\left|\gamma\right|}e^{\gamma^{2}/2R^{2}}\left[e^{-a\left|\gamma\right|}\erfc\left(\frac{1}{\sqrt{2}}\left(\frac{\left|\gamma\right|}{R}-aR\right)\right)+e^{a\left|\gamma\right|}\erfc\left(\frac{1}{\sqrt{2}}\left(\frac{\left|\gamma\right|}{R}+aR\right)\right)\right].
\end{eqnarray}
Consequently, we have 
\[
\int_{-\infty}^{\infty}\frac{e^{\pm i\sin\theta k_{0}x}e^{-x^{2}/2R^{2}}}{1+i\alpha k_{0}x}dx=\pi\left|\gamma\right|e^{\frac{\gamma{}^{2}}{2R^{2}}}e^{\pm\mbox{sign}\left(\gamma\right)a\left|\gamma\right|}\mbox{Erfc}\left[\frac{1}{\sqrt{2}}\left(\frac{\left|\gamma\right|}{R_{x}}\pm\mbox{sign}\left(\gamma\right)aR_{x}\right)\right].
\]
This leads to expression \eqref{eq:skSS}, since the $\pm$ can be
replaced by $+$ by extending the angle range to interval $0\leq\theta\leq2\pi$.

The double scattering field is given by the expression 
\begin{equation}
E^{(2)}(\mathbf{k})=\frac{\hbar\Gamma}{id}\frac{e^{ik_0r}}{r}\int d^{3}\mathbf{r}\rho(\mathbf{r})e^{-i\mathbf{k}\cdot\mathbf{r}}\int d^{3}\mathbf{r'}\rho(\mathbf{r'})K(\mathbf{r},\mathbf{r}')\beta^{(1)}(\mathbf{r}').\label{double sf}
\end{equation}
These integrals can be decoupled for a Gaussian density distribution,
using the following change of variables 
\begin{flalign}
\mathbf{u}=\frac{\mathbf{r}-\mathbf{r}'}{\sqrt{2}},\qquad & \mathbf{w}=\frac{\mathbf{r}+\mathbf{r}'}{\sqrt{2}},
\end{flalign}
so that $\rho(\mathbf{r})\rho(\mathbf{r}')=\rho(\mathbf{u})\rho(\mathbf{w})$,
and assuming that the quadratic terms in $u^{2}$ and $v^{2}$ can
be neglected in the Zeeman term (approximation of small frequency
shift $2\alpha^{2}\sigma^{2}\ll1$). Then we obtain: 
\begin{equation}
E^{(2)}(\mathbf{k})=E_{0}\frac{e^{ik_0r}}{r}\int d^{3}\mathbf{u}\rho(u)G(\sqrt{2}u)e^{-i\mathbf{u}\cdot(\mathbf{k}_{0}+\mathbf{k})/\sqrt{2}}\int\frac{d^{3}\mathbf{w}\rho(w)e^{i\mathbf{w}\cdot(\mathbf{k}_{0}-\mathbf{k})/\sqrt{2}}}{1+i\sqrt{2}\alpha\mathbf{w}.\hat{x}}.\label{eq:S2}
\end{equation}
Similarly to the single scattering case, the second integral $I_{w}$
in \eqref{eq:S2} leads to 
\begin{equation}
I_{w}(\theta)=\frac{\sqrt{\pi}}{2}\frac{N}{\left|\alpha\right|\sigma}\exp\left(\frac{1}{4\alpha^{2}\sigma^{2}}+\frac{\mbox{sign}\left(\alpha\right)\sin\theta}{2\left|\alpha\right|}-\frac{\sigma^{2}}{4}(1-\cos\theta)^{2}\right)\erfc\left(\frac{1}{2}\left(\frac{1}{\left|\alpha\right|\sigma}+\mbox{sign}\left(\alpha\right)\sigma\sin\theta\right)\right).
\end{equation}
The first integral, $I_{u}$, in \eqref{eq:S2} is calculated using
spherical coordinates $(u,\theta_{u},\phi_{u})$. The integral over
$\phi_{u}$ leads to 
\begin{eqnarray}
I_{u}(\theta) & = & \frac{N}{i2\sqrt{\pi}k_{0}R^{3}}\int_{0}^{\infty}ue^{-u^{2}/2R^{2}+i\sqrt{2}k_{0}u}du\int_{0}^{\pi}\sin\theta_{u}e^{-i\frac{k_{0}u}{\sqrt{2}}(1+\cos\theta)\cos\theta_{u}}J_{0}\left(\frac{k_{0}u\sin\theta\sin\theta_{u}}{\sqrt{2}}\right)d\theta_{u}\\
 & = & \frac{N}{i\sqrt{2\pi}k_{0}^{2}R^{3}\cos\frac{\theta}{2}}\int_{0}^{\infty}e^{-u^{2}/2R^{2}+i\sqrt{2}k_{0}u}\sin\left(\sqrt{2}k_{0}u\cos\frac{\theta}{2}\right)du\nonumber \\
 & = & -\frac{b_{0}}{8\cos\frac{\theta}{2}}\left[\exp\left(-\sigma^{2}(1+\cos\frac{\theta}{2})^{2}\right)\erfc\left(-i\sigma(1+\cos\frac{\theta}{2})\right)-\exp\left(-\sigma^{2}(1-\cos\frac{\theta}{2})^{2}\right)\erfc\left(-i\sigma(1-\cos\frac{\theta}{2})\right)\right],\nonumber 
\end{eqnarray}
where we have introduced the optical thickness $b_{0}=2N/\sigma^{2}$.

\end{widetext}


\end{document}